\documentclass[structuredabstract]{aa}

\newcommand{\emm}[1]{\ensuremath{#1}}   
\newcommand{\emr}[1]{\emm{\mathrm{#1}}} 
\newcommand{\unit}[1]{\emm{\, \emr{#1}}}
\newcommand{\wmqsr}{W\,m$^{-2}$\,sr$^{-1}$\xspace}
\newcommand{\mum} {\unit{\mu m}\xspace}
\newcommand{\K}   {\unit{K}\xspace}
\newcommand{\cmc}{\unit{cm^{-3}}\xspace}
\newcommand{\ex}[2]{\ensuremath{#1 \times 10^{#2}}\xspace}
\newcommand{\spitzer}{\textit{Spitzer}\xspace}
\newcommand{\HH}    {\mbox{H$_2$}\xspace}           
\newcommand{\pah}    {\mbox{PAH$^0$}\xspace}           
\newcommand{\pahp} {\mbox{PAH$^+$}\xspace} 

\usepackage{subcaption}
\usepackage{graphicx}
\usepackage{natbib}
\usepackage{xspace}
\xspaceaddexceptions{]\}}
\usepackage{ulem}

\usepackage[update,prepend]{epstopdf}
\usepackage{booktabs}
\usepackage{amsmath}
\usepackage{color}

\usepackage{memhfixc}
\usepackage{txfonts}
\usepackage{multirow}

\usepackage[colorlinks, 
        linkcolor = blue, 
        citecolor = blue,
        breaklinks = true]{hyperref}

\usepackage{color}

\begin{document}

\title{Mixed aliphatic and aromatic composition of evaporating very small grains in NGC 7023 revealed by the 3.4/3.3 \mum ratio}

\author{
P.~Pilleri\inst{1,2}, 
C.~Joblin\inst{1,2}, 
F.~Boulanger\inst{3},
T.~Onaka\inst{4}
}
 \institute{%
Universit\'e de Toulouse; UPS-OMP; IRAP;  Toulouse, France
\and
CNRS; IRAP; 9 Av. colonel Roche, BP 44346, F-31028 Toulouse cedex 4, France 
\and 
Institut d'Astrophysique Spatiale, 91405, Orsay, France
\and
Department of Astronomy, Graduate School of Science, University of Tokyo, Tokyo 113-0033, Japan
}

\date{Received ? accepted?}

\abstract
{A chemical scenario was proposed for photon-dominated regions (PDRs) according to which UV photons from nearby stars lead to the evaporation of very small grains (VSGs) and the production of gas-phase polycyclic aromatic hydrocarbons (PAHs).}
{Our goal is to achieve better insight into the composition and evolution of evaporating very small grains (eVSGs) and PAHs through analyzing the infrared (IR) aliphatic and aromatic emission bands.}
{We combined spectro-imagery  in the near- and mid-IR to study the spatial evolution of the emission bands in the prototypical PDR NGC~7023. We used near-IR spectra obtained with the IRC instrument onboard AKARI to trace the evolution of the 3.3\mum and 3.4\mum bands, which are associated with aromatic and aliphatic C$-$H bonds on PAHs.  The spectral fitting involved an additional broad feature centered at 3.45\mum that is often referred to as the plateau. Mid-IR observations obtained with the IRS instrument onboard the \spitzer Space Telescope were used to distinguish the signatures of eVSGs and neutral and cationic PAHs.  We  correlated the spatial evolution of all these bands with the intensity of the UV field given in units of the Habing field $G_0$ to explore how their carriers are processed.  }
{The intensity of the 3.45\mum plateau shows an excellent correlation with that of the 3.3\mum aromatic band (correlation coefficient R = 0.95) and a relatively poor correlation with the aliphatic 3.4\mum band (R=0.77). This indicates that the 3.45\mum  feature is dominated by the emission from aromatic bonds. We show that the ratio of the 3.4\mum and 3.3\mum band intensity ($I_{3.4}/I_{3.3}$)  decreases by a factor of 4 at the PDR interface from the more UV-shielded layers ($G_0 \sim 150, I_{3.4}/I_{3.3} = 0.13$) to the more exposed layers ($G_0 > \ex{1}{4}, I_{3.4}/I_{3.3} = 0.03$).  The intensity of the 3.3\mum band relative to the total neutral PAH intensity shows an overall increase with $G_0$, associated with an increase of both the hardness of the UV field and the H abundance. In contrast, the intensity of the 3.4\mum band relative to the total neutral PAH intensity decreases with $G_0$, showing that their carriers are actively destroyed by UV irradiation and are not efficiently regenerated. The transition region between the aliphatic and aromatic material is found to correspond spatially with the transition zone between neutral PAHs and eVSGs.}
{We conclude that the photo-processing of eVSGs leads to the production of PAHs with attached aliphatic sidegroups that are revealed by  the 3.4\mum emission band. Our analysis provides evidence for the presence of very small grains of mixed aromatic
and aliphatic composition in PDRs.    }

\date{Received ?; Accepted ?}

\authorrunning{Pilleri P., et al.}
\titlerunning{The mixed aliphatic/aromatic composition of eVSGs in NGC 7023  revealed by the 3.4/3.3 \mum ratio}

\maketitle

\section{Introduction}

A significant fraction of interstellar carbon \citep[up to 20\%,][]{joblin92, tielens05} is tied up in the carriers of the aromatic infrared bands (AIBs). The most intense AIBs are observed at 3.3, 6.2, 7.7, 8.6, 11.3, and 12.7\mum and are 
 generally attributed to stochastically heated polycyclic aromatic hydrocarbons  \citep[PAHs, ][]{leger84, allamandola85}. 
 This set of emission bands shows systematic variations in the relative intensity and central wavelength of the bands \citep{peeters02}, reflecting variations in the local physical conditions and a chemical evolution of their carriers. 
From the analysis of spectro-imagery data on several Galactic photon-dominated regions (PDRs),  the  variations of the mid-IR ($5.5-15\mum$) bands were attributed  to the emission of three different populations: PAH cations and neutrals (\pahp and \pah) and evaporating very small grains \citep[eVSGs, ][]{rapacioli05, berne07, pilleri12}. A chemical scenario has been proposed according
to which eVSGs evaporate under UV irradiation, producing free-flying PAHs. Candidates such as PAH clusters  \citep{rapacioli06, montillaud14} or Fe-PAH complexes \citep{simon09}  have been proposed as models for these eVSGs. \citet{jones12a} recently proposed a model for the evolution of carbonaceous dust in the interstellar medium in which the smallest particles are aromatized, whereas the largest ones remain H-rich and predominantly aliphatic. Such a scenario would reconcile the observation of both the aromatic emission bands in UV-irradiated regions and the 3.4\mum CH aliphatic absorption feature as a main feature of interstellar dust in galaxies \citep{dartois04}. To develop this evolutionary scenario further,  we report here a study of the 3.4\mum emission band, which is observed as a minor band at the side of the main AIB at 3.3\mum. Characteristic of aliphatic CH bonds, this band has been attributed to methyl sidegroups attached to PAHs  \citep{muizon90} or to superhydrogenated PAHs \citep{bernstein96}. The observed 3.4/3.3\,\mum intensity ratio ($I_{3.4}/I_{3.3}$) is found to decrease  for an increasing intensity of the UV field \citep{geballe89, joblin96, sloan97,mori14}, reflecting the photo-destruction of the more fragile bonds linked to the 3.4\mum band. 

\begin{figure*}[t]      
\centering
\includegraphics[trim = 0cm 0cm 0cm 0cm, clip, width=1\hsize]{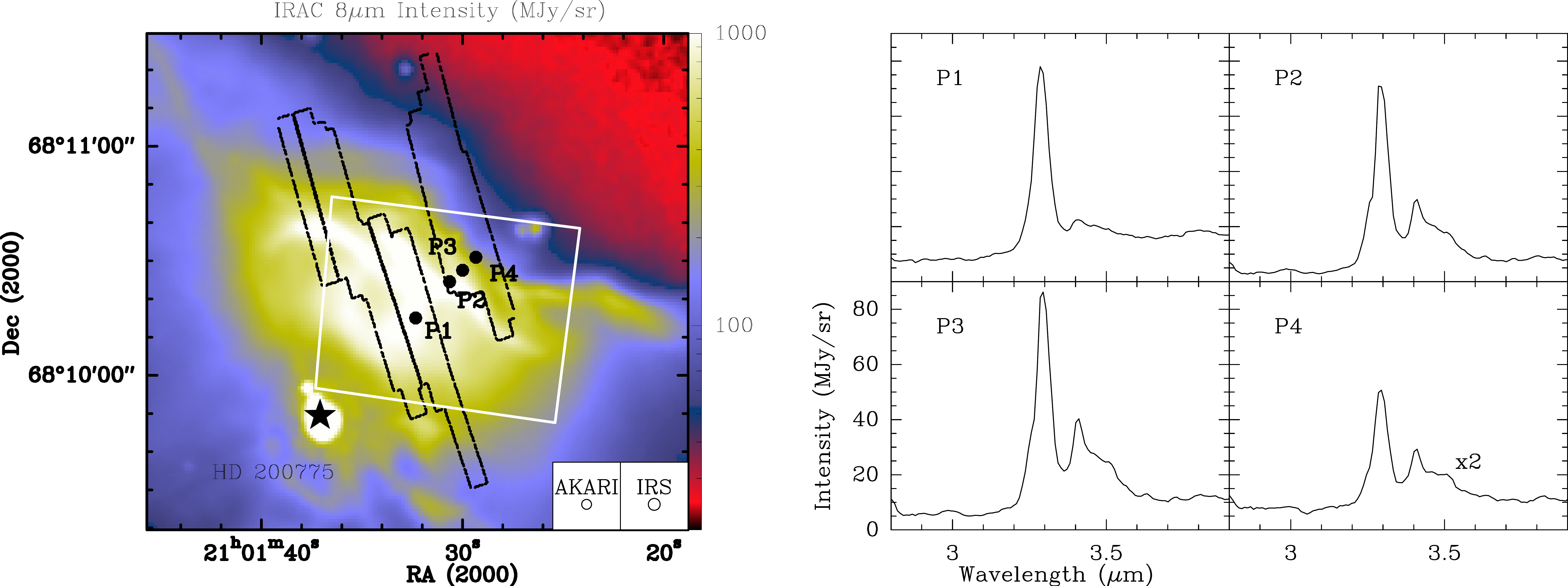}
\caption{In color scale, the \spitzer IRAC 8\,\mum image of the NGC 7023 NW PDR \citep{werner04}.  The dark lines demarcate the region in which AKARI-IRC data were obtained, the white rectangle represents the \spitzer-IRS field of view. The star represents the position of HD 200775.  The insets represent the AKARI and IRS beam sizes. AKARI spectra between 2.8 and 3.9\mum are shown to the right. They correspond to the positions P1-P4 that are marked as black dots on the map. } 
\label{fig_summary}
\end{figure*}

In this paper we present   2D spectroscopy  of  the PDR associated with the reflection nebula NGC 7023 NW   obtained with the AKARI and \spitzer space telescopes. The combination of these datasets  allows us to study the spatial variations of the 3.3 and 3.4 $\mu$m emissions due to aromatic and aliphatic bonds with those of the \pah and eVSG emissions. 
 In Sect. 2 we present the observations and the data reduction. In Sect. 3  we describe the data analysis tools and present the observational results. We discuss these results in Sect. 4. Finally, the conclusions are presented in Sect. 5.

\section{Observations}

\subsection{NGC 7023 NW} 

NGC 7023 is a reflection nebula in the Cepheus constellation illuminated by the B2-5Ve binary star HD~200775 \citep{alecian08}. The distance of  HD~200775 to the Sun measured by Hipparcos  is 430$^{+160}_{-90}$\,pc, but a recent analysis revised the distance to $320\pm51$pc  \citep{benisty13}. 
  The nebula has been shaped by the star formation process, which led to the formation of a cavity \citep{fuente93}. The interaction of the UV photons with the walls of the parent molecular cloud has produced several PDRs to the northwest, south, and east of the star. The brightest PDR is found about 45\arcsec\, NW from the binary star (hereafter the NW~PDR, or NGC~7023~NW). \citet{pilleri12} estimated the intensity of the radiation field at the PDR front  to be G$_0 \sim 2600$ in units of the Habing field  \citep{habing68}.
Previous observations of different gas lines have shown that this region hosts structures at different gas densities: $n_{\rm H}\sim100$\,\cmc\ in the cavity \citep{berne12},  $10^5-10^6$\,\cmc\ in the filaments that are observed in the mm \citep{fuente96} and near-IR \citep{lemaire96, martini97}, and $\sim10^4$\cmc\ in the molecular cloud \citep{gerin98}. 

\begin{figure*}[ht]     
\centering
\includegraphics[angle = -90, trim = 0cm 0cm 0cm 0cm, clip, width=0.95\hsize]{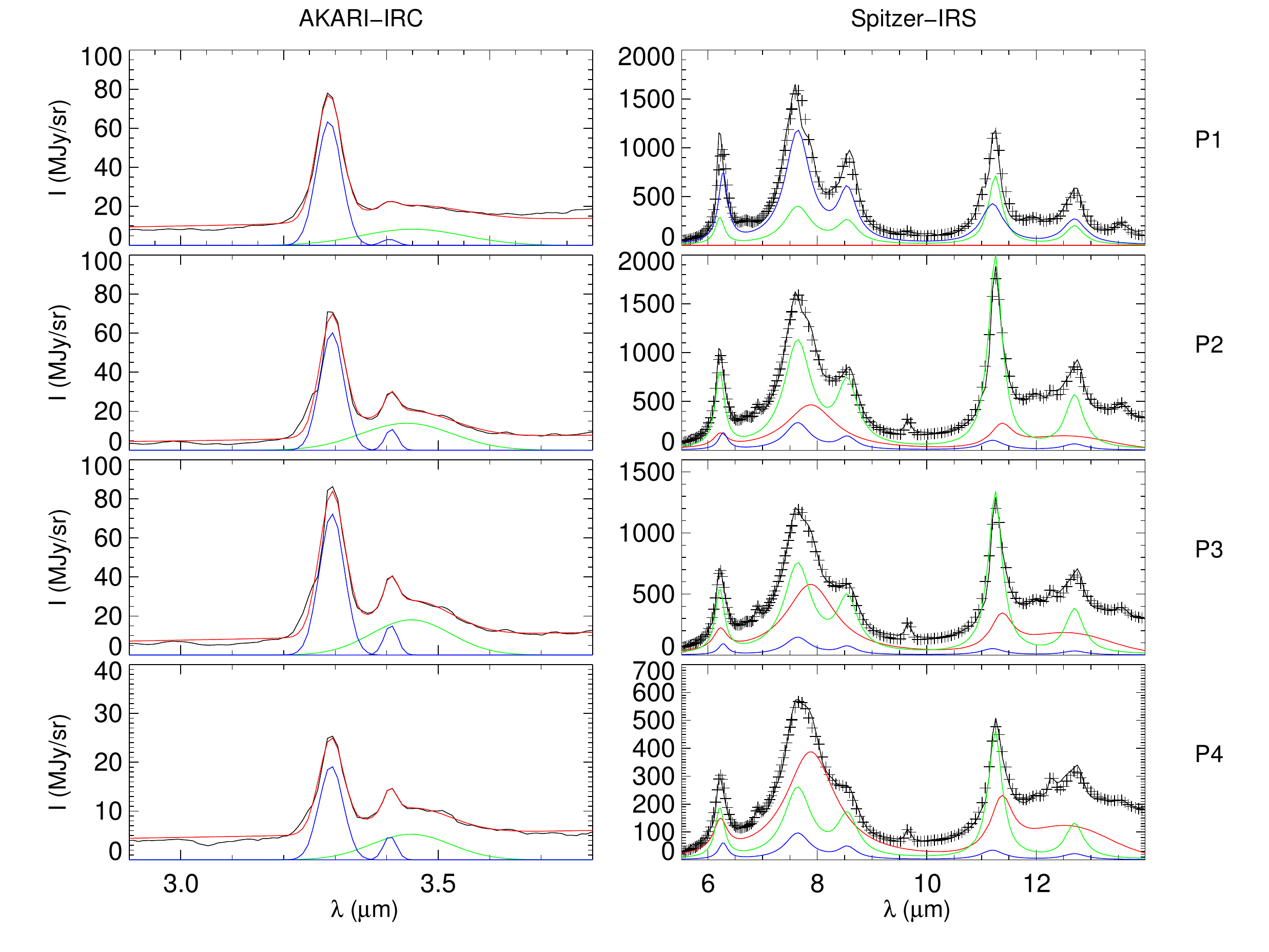}
\caption{Summary of the observations and spectral analysis for positions P1 to P4 (top to bottom row). {\it First column:} the observed spectrum (black) and its fit (red) for the AKARI observations. The fit of the 3.29 and 3.40 \mum features are displayed in blue, and the 3.45 \mum plateau is shown in green.  {\it Second column:} observed Spitzer-IRS spectrum (crosses), its fit (black) and the decomposition obtained with PAHTAT (red: eVSG, green: \pah, blue: \pahp).}\label{obs}
\end{figure*}

\subsection{Near- and mid-IR observations with \spitzer and AKARI}
\newcommand{\ratioIRnear}{$I_{3.4}/I_{3.3}$\xspace}
\label{nIR_obs}

The infrared camera \citep[IRC,][]{onaka07} onboard the AKARI space telescope \citep{murakami07} was used  to obtain near-IR (2.5-5\mum) spectro-imagery of NGC 7023 NW. The observations were performed as part of the open-time program NESID 
of the AKARI post-helium phase between  October 2008 and  January 2010.
They were carried out using the grism spectroscopy mode  with the narrow Nh slit, which provides a spectral resolution of about 0.02\,$\mu$m and a spatial resolution of  $\sim3\arcsec$  \citep{onaka07}.
The slit position was slightly shifted for the different runs to make dithering observations.
In each pointed observation, eight to nine exposure cycles in spectroscopy were obtained together with one exposure in imaging in the N3 
(3.2\,$\mu$m) band, which gives the accurate position information.  The on-source
integration time for each observation is typically 360\,s.
The data were processed with the latest version of the spectroscopic toolkit
(version 20110301) for the phase 3 observation\footnote{\url{http://www.ir.isas.jaxa.jp/ASTRO-F/Observation/DataReduction/IRC/}}.
The NGC 7023 NW observations  consisted  of 20 stripes of length $\sim50\arcsec$,  each with a slightly different position and orientation. 
The stripes were then combined in a mosaic using the software {\it montage}\footnote{\url{http://montage.ipac.caltech.edu}}.  The field of view of the final mosaic  results from the overlap of the slit coverages of all the stripes and is displayed in Fig. \ref{fig_summary}. The near-IR features vary smoothly between nearby positions but are significantly different with distance to the star, as shown by the four representative spectra in Fig. \ref{fig_summary} that correspond to offsets of P1  [-29\arcsec, 24\arcsec]; P2 [-36\arcsec, 35\arcsec], P3 [-39\arcsec, 38\arcsec], P4 [-43\arcsec, 41\arcsec] relative to the position of HD 200775.

We compare this dataset with the \spitzer-IRS observations of NGC 7023, which provides spectral imagery in the wavelength range [5.5-14]\mum and covers the whole filamentary region of the NW PDR with a spatial resolution of 3.6\arcsec (see Fig.  \ref{fig_summary}). This dataset has been described in detail in previous studies,
for example,{\it } \citet{werner04}, \citet{berne07}, and \citet{pilleri12}.

\section{Spectral analysis and spatial distributions}

\subsection{Near-infrared}

The first column of Fig. \ref{obs} shows the AKARI-IRC spectra observed at the positions P1-P4. The spectra are dominated by the two emission bands that peak at 3.29 and 3.40\mum, in addition to an underlying plateau. 
To fit these spectra we used two narrow Gaussians for the 3.29 and 3.40\mum features and a broader Gaussian centered at 3.45\mum to account for the plateau. We let the central wavelength of the Gaussians vary by 0.01\mum to account for small shifts and treated the band widths and intensities as free parameters. We simultaneously fit the observed spectrum using a linear combination of these features and a linear continuum. 
The resulting full widths at half maximum of the bands were found to be $\sim 0.03$\,\mum for the 3.29 and 3.40 \mum bands and $\sim 0.1$\,\mum for the 3.45 \mum plateau. As shown in Fig.\,\ref{obs}, this yields very good fits of the spectra, even in cases with a strong plateau and faint superposed 3.40\mum band.

\begin{figure*}[ht]     
\centering
\includegraphics[trim = 0cm 1cm 0cm 0cm, clip,angle=-90, width=0.75\hsize]{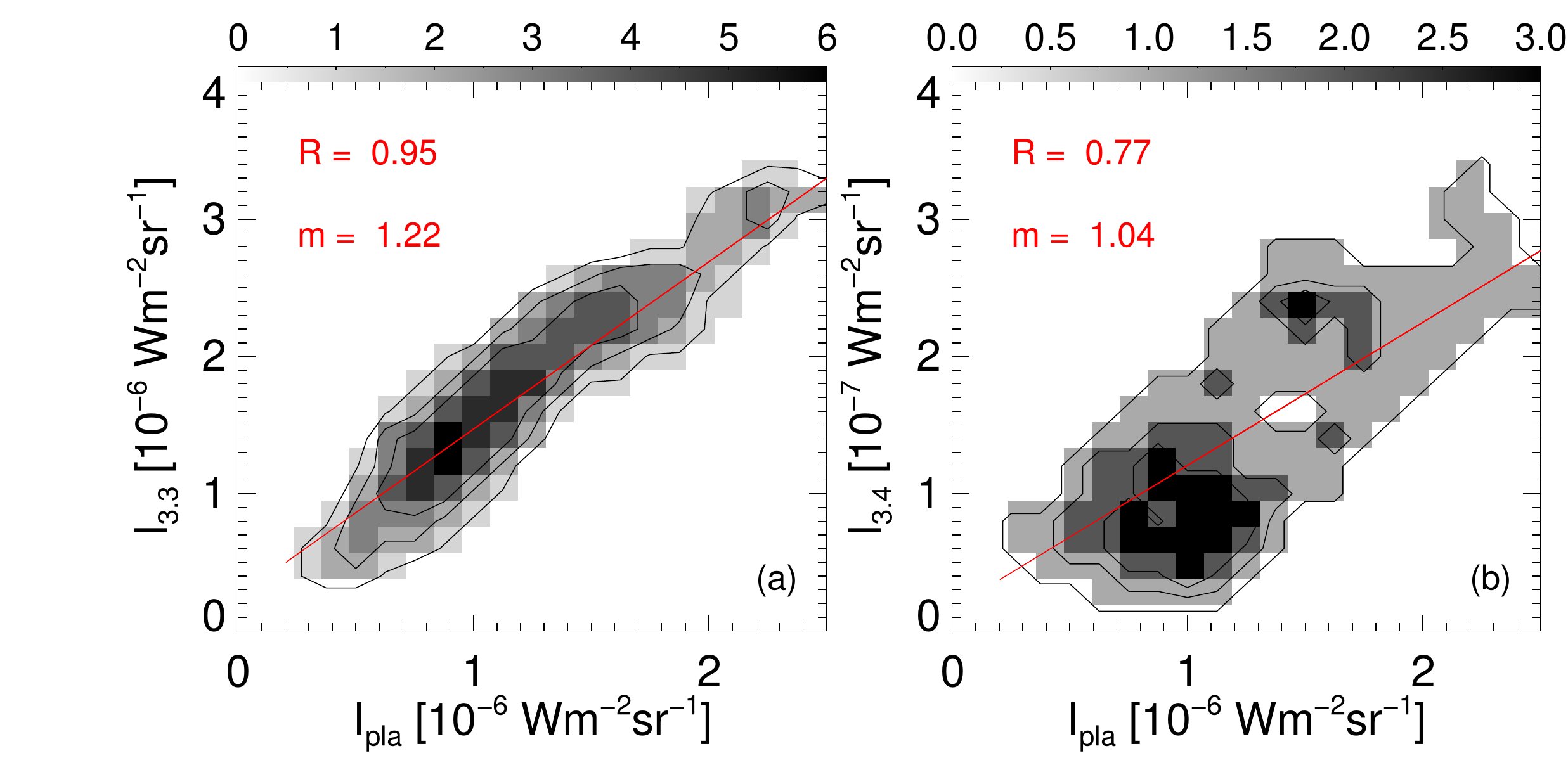}
\caption{Pixel-to-pixel correlations of the integrated intensity of the 3.3\mum band ($I_{3.3}$), 3.4\mum band ($I_{3.4}$), and the 3.45\mum plateau ($I_{pla}$). The correlation coefficient $R$ and the slope $m$ of the linear fit to the points are also given. $I_{pla}$ correlates better with the intensity of the aromatic 3.3\mum band than with the aliphatic 3.4\mum band. This indicates that the plateau is  dominated by emission from aromatic bonds. }\label{correlations_patate}
\end{figure*}

The 3.29\mum band is commonly attributed to the C$-$H in-plane stretching mode in PAHs, whereas the 3.40\mum band and the  plateau are more difficult to attribute unequivocally. \citet{steglich13} concluded that it is difficult from a spectroscopic point of view to favor the attribution of the 3.4\mum band either to methyl side groups attached to PAHs or to superhydrogenated PAHs. Still, \citet{joblin96} showed that the evolution of the 3.4/3.3\mum band ratio with the UV field can be explained by methylated PAHs. 
The 3.45\mum plateau is most likely the superposition of a number of faint features including other aliphatic modes, hot band emission of aromatic CH \citep{barker87} and combination bands of C$-$C and C$-$H aromatic modes \citep{allamandola89}. In Fig.\,\ref{correlations_patate} we show the pixel-to-pixel correlation of the integrated intensity of the 3.45\mum plateau ($I_{pla}$) with that of the 3.29\mum  ($I_{3.3}$) and 3.40\mum  ($I_{3.4}$) bands. $I_{pla}$ correlates better with $I_{3.3}$  (correlation coefficient $R = 0.95$) than with $I_{3.4}$ ($R = 0.77$). This  suggests that the plateau is dominated by the emission from aromatics (hot bands and combination bands). Since  our goal is to study the behavior of the aromatic vs aliphatic evolution, we exclude the plateau from the analysis in the following.

\subsection{Mid-infrared}

To extract the emission of the different AIB carriers, we used the results of the PAHTAT procedure \citep{pilleri12}.  Figure \ref{obs} shows the mid-IR spectra and the corresponding fits for the four positions P1-P4.   PAHTAT also allows estimating the value of  $A_V$ along the line of sight assuming that the AIB carriers and other dust populations are well mixed. From these $A_V$ values, we reconstructed the deredenned mid-IR and near-IR spectra by using the extinction curve from \citet{weingartner01} for an $R_V$ value of 5.6 \citep{witt06}. These unreddened spectra are used in the following analysis.  The 3.3 and 3.4\mum bands are sufficiently close in wavelength for their ratio is to be
not significantly affected by the extinction correction.

\begin{figure}[ht]      
\centering
\includegraphics[trim = 0cm 0cm 0cm 0cm, clip, width=0.7\hsize]{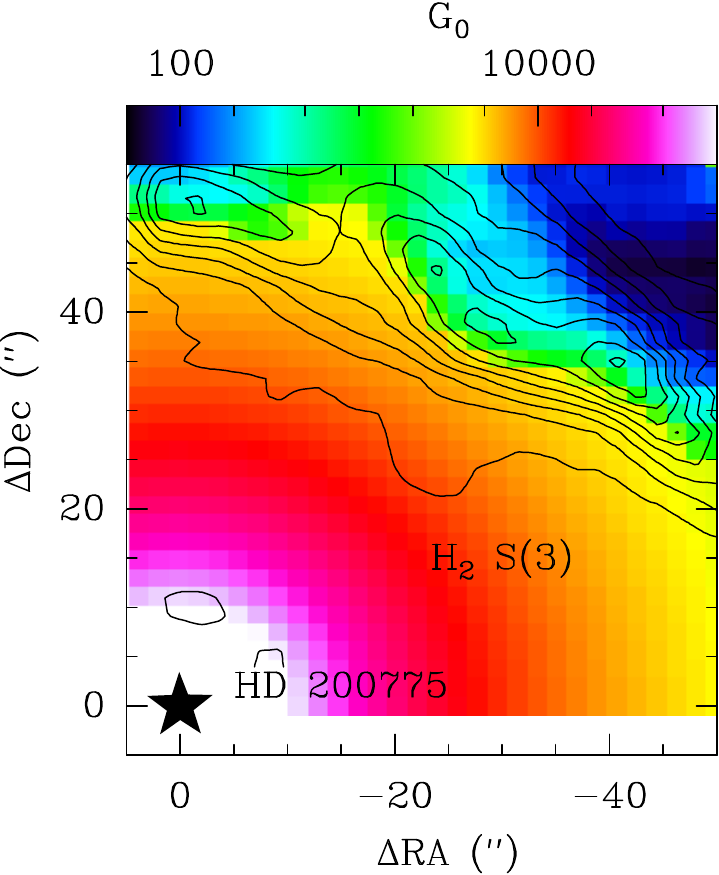}
\caption{Map of G$_0$ in NGC 7023 NW  (color scale). The H$_2$ S(3) integrated intensity is shown for reference (linear steps of \ex{1.7}{-7}\,W\,m$^{-2}$\,sr$^{-1}$). The values of G$_0$ between the star and the PDR were obtained assuming geometrical dilution of the stellar radiation field and a local extinction value of $A_V = 1.5$.  Inside the PDR, we used the fraction of carbon locked in eVSG as a probe for G$_0$ \citep{pilleri12}.}\label{fig_g0}
\end{figure}

 Table \ref{table_resume} shows the relevant results of the fit for the positions P1-P4, that is, the integrated intensity of the near-IR bands, of the \pahp, \pah, eVSG, and of the total AIBs, as well as  the fraction of carbon locked in eVSGs ($f_{\rm eVSG}$).   In \citet{pilleri12}, we have shown that $f_{eVSG}$ can be derived from every pixel and then used as a probe for $G_0$  in the region of the PDR with significant  eVSG emission.  For NGC 7023 NW, this region starts at a distance of $\sim 45\arcsec$ from HD~200775, corresponding to $G_0 = 2600$. In the more exposed layers of the PDR, we obtain an estimate of the UV  field intensity by assuming geometrical dilution of the stellar radiation field,  a stellar temperature of $T_{eff} = 15\,000$\K \citep{finkenzeller85}, and a local extinction of $A_V = 1.5$ \citep{pilleri12}.  The total map of $G_0$ we obtain is shown in Fig. \ref{fig_g0}.

\begin{table}[htdp]
\caption{Results of the fit of the mid- and near-IR observations.}
\begin{center}
\begin{tabular}{c|c|cccccc}
\toprule
                        &               Unit                            &       P1                      &               P2              &               P3                      &                 P4              \\              
\midrule
$I_{eVSG}$      &               $10^{-5}$\,\wmqsr       &       0.00                    &                 4.81    &               6.00                    &               4.01    \\              
$I_{PAH^0}$     &               $10^{-5}$\,\wmqsr       &       4.04                    &                 11.4    &                7.60           &               2.63            \\              
$I_{PAH^+}$     &               $10^{-5}$\,\wmqsr       &       8.87                    &                 2.13            &                1.10   &               0.73            \\              
$I_{AIB}$               &               $10^{-5}$\,\wmqsr       &       12.9                    &                  18.3           &               14.7                    &               7.36            \\              
\midrule
$A_V$           &               mag                             &       $<2$                    &               15              &               19                      &               23              \\              
$f_{eVSG}       $       &               -                               &       0.00                    &               0.42            &               0.58                    &               0.71            \\              
$G_0$           &               -                               &       7000    &               2600            &               200                     &                 150             \\              
\midrule
$I_{3.3}$               &               $10^{-7}$\,\wmqsr       &       11.0                    &       15.8            &               22.3                         &               6.47 \\         
$I_{3.4}$               &               $10^{-7}$\,\wmqsr       &       0.31                    &       1.42                    &                 2.34                    &               0.80\\          
$I_{pla}$               &               $10^{-7}$\,\wmqsr       &       5.36                    &       13.5            &               15.6                    &                4.80     \\             
$I_{3.4}/I_{3.3}$&                      -                       &       0.028           &       0.090           &                 0.11                    &               0.13    \\
\bottomrule
\end{tabular}
\end{center}
\label{table_resume}
\end{table}%

\begin{figure}[t]       
\centering
\includegraphics[width=0.9\hsize]{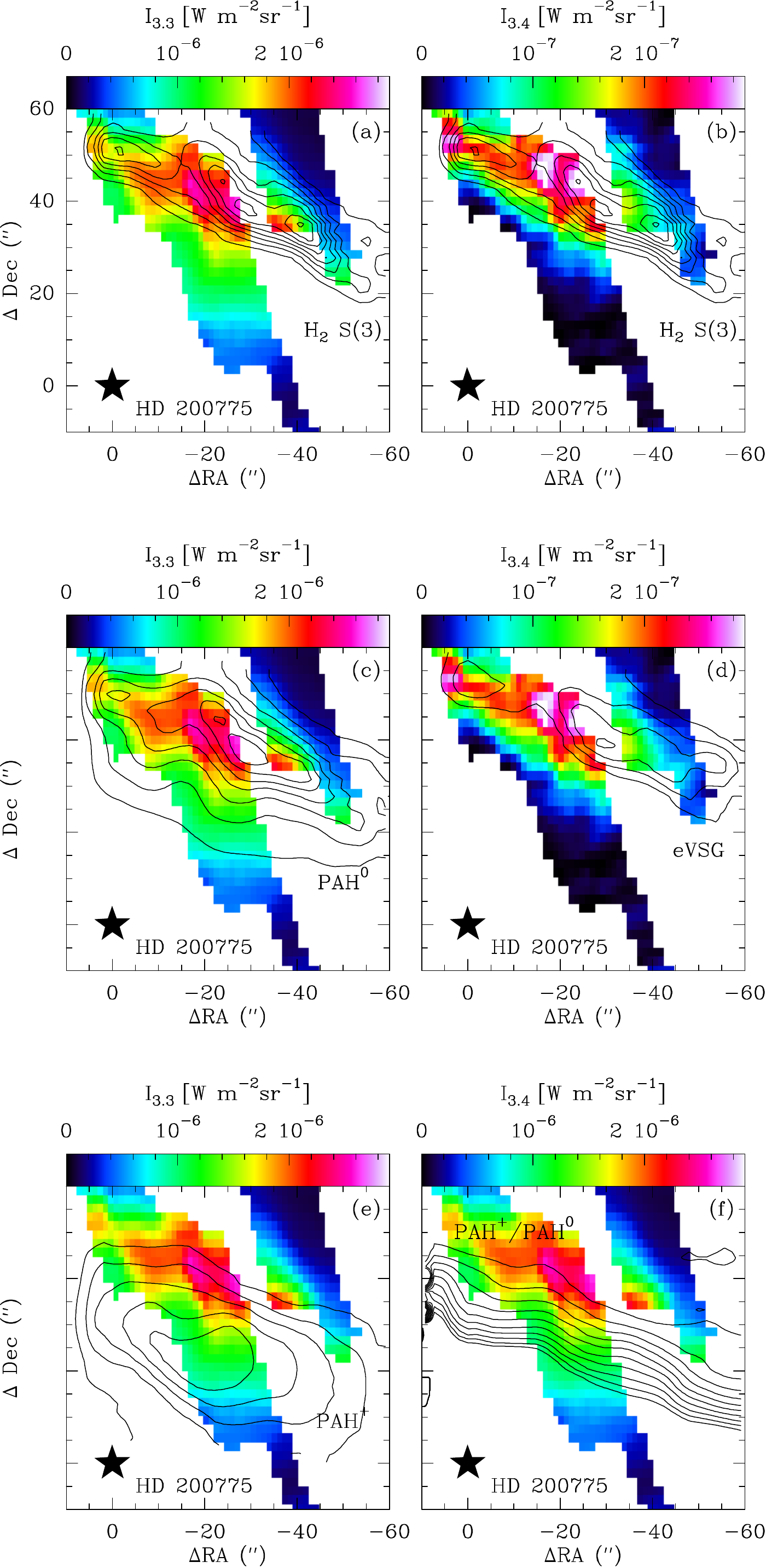}
\caption{{\it Top:} Spatial distribution of the 3.3\mum (a) and 3.4\mum (b) integrated intensity as observed by the IRC instrument (color scale).  
The contours represent the integrated intensity of the \HH S(3) line at 9.7\mum (linear steps of \ex{1.7}{-7}\,W\,m$^{-2}$\,sr$^{-1}$), and the star indicates the position of HD 200775. {\it Middle:} Comparison of the integrated intensity of the bands at 3.3 and 3.4\mum (color scale) with the spatial distribution of the  \pah and eVSG intensity, respectively (contours steps of \ex{1.7}{-5}\,W\,m$^2$\,sr$^{-1}$,  starting at \ex{1.7}{-5}\,W\,m$^2$\,sr$^{-1}$).  {\it Bottom:}   Comparison of the integrated intensity of the 3.3\mum band with that of \pahp and with the ratio $I_{PAH^+}/I_{PAH^0}$. In (f) the contours are shown in steps of 0.3, with increasing values while approaching the star from the PDR.}\label{figure_pahtat}
\end{figure}

\subsection{Spatial distributions}

 \begin{figure*}[ht]    
\centering
\includegraphics[angle = -90, trim = 0cm 0cm 0cm 0cm, clip, width=\hsize]{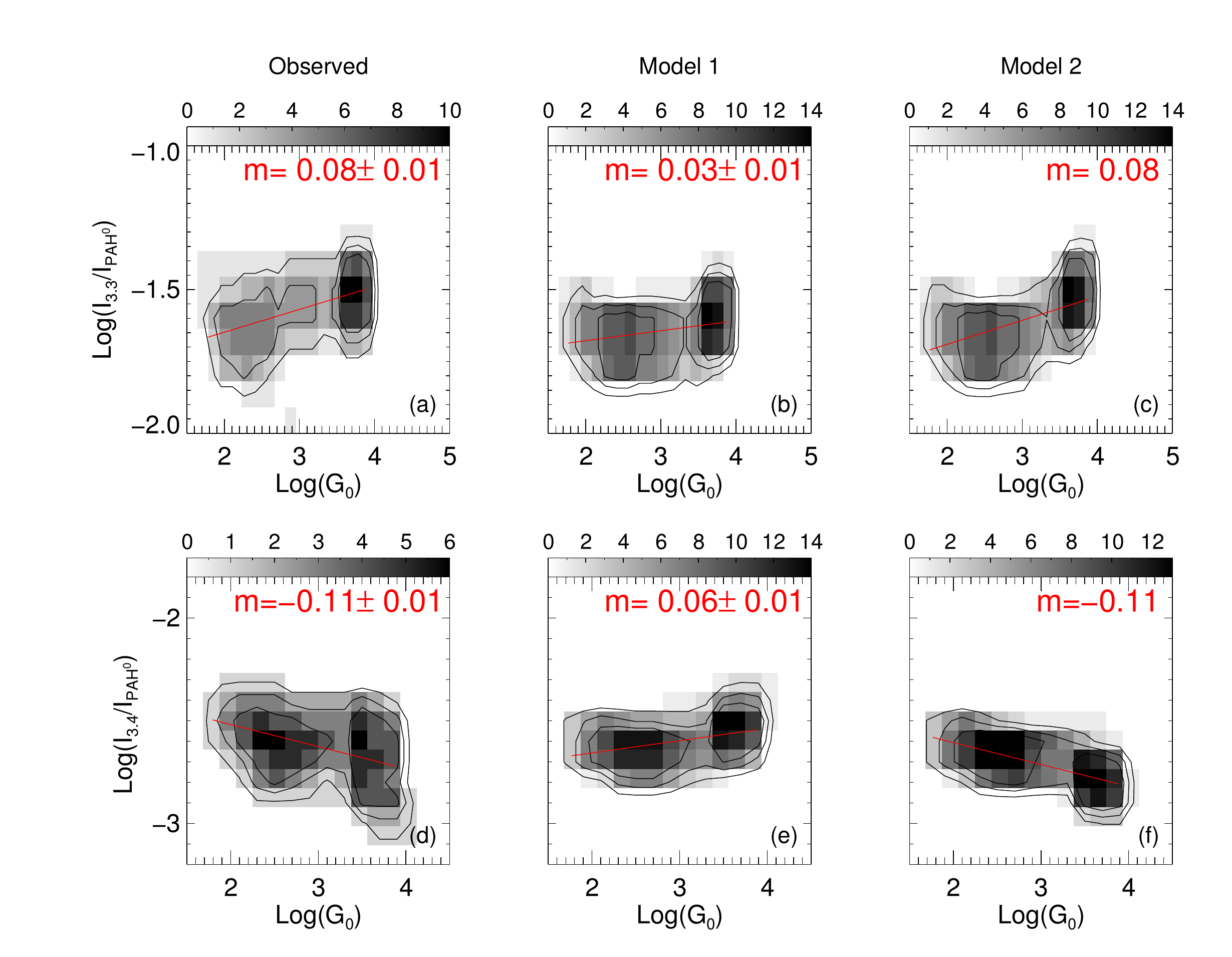}
\caption{{\it First column:} Correlation between the observed $I_{3.3}/I_{PAH0}$ and  $I_{3.4}/I_{PAH0}$ as a function of $G_0$.  {\it Second column:}  Same plots for  model maps of $I_{3.3}/I_{PAH0}$ and  $I_{3.4}/I_{PAH0}$  computed using Eqs. 6 and 8 and  the obtained values for $I_{PAH^+}/I_{PAH^0}$ (see text for details). 
 {\it Third column:} Semi-empirical model that in addition includes the dependency with $G_0$ on the intensity at 3.3 and 3.4 \mum relative to all \pah bands (see text).  In all panels, the linear fit to the data points (red line) is shown as well as its slope $m$.}
\label{figure_correlations}
\end{figure*}

Applied to the full IRS data cube, PAHTAT allows deriving the spatial distribution of the intensity of  \pahp, \pah  , and eVSGs (top panel in Fig. \ref{figure_pahtat}). These results have been described in detail in \citet{pilleri12} and can be summarized as follows: \pahp dominate the mid-IR emission between the illuminating star and the PDR front; \pah are the most abundant species at the PDR front as traced by the emission of the H$_2$ rotational lines; eVSGs dominate the emission slightly deeper inside the cloud; the transition  between \pah and eVSGs corresponds to the peak of the H$_2$ S(3) line, which is found at a magnitude of visual extinction, $A_{\rm V} \sim 1$ \citep{pilleri12}.

Figures \ref{figure_pahtat}a,b show the integrated intensity maps of the 3.3\mum and 3.4\mum bands. Both  maps peak toward the H$_2$ S(3) filament that delineates the PDR front. The 3.3\mum band is also detected in the cavity between the star and the PDR, whereas the 3.4\mum band is detected only toward the PDR.  Neither the 3.3\mum nor the 3.4\mum bands is detected in the more shielded layers of the PDR,  farther away from HD~200775 than the H$_2$ filaments. 

 In Fig. \ref{figure_pahtat}c,d we compare these maps with the spatial distribution of \pah and eVSGs intensities obtained with PAHTAT.  A very good spatial correlation is found between $I_{3.3}$ and $I_{PAH^0}$ (Fig.  \ref{figure_pahtat}c). In contrast, the 3.4\mum band is more localized on the dense PDR with a good spatial correlation between $I_{3.4}$ and $I_{eVSG}$ (Fig.  \ref{figure_pahtat}d).
Still, the 3.3\mum band is observed inside the cavity, which indicates that there is a contribution from \pahp to its emission, and/or that a fraction of \pah is present in the cavity (Fig.\ref{figure_pahtat}e).  The intensity at 3.3\mum  decreases by a factor $> 5$ (from \ex{3}{-6} to \ex{6}{-7} W\,m$^{-2}$\,sr$^{-1}$) in regions where  $I_{PAH^+}/I_{PAH^0} \gtrsim 3$ (Fig.\,\ref{figure_pahtat}f).   Since we are interested in the evolution of the 3.4\mum band relative to the 3.3\mum band, we restrain our analysis below
to the PDR front and mask the pixels that have $I_{PAH^+}/I_{PAH^0} >       1$ and those where the 3.4\mum band is not detected.

\section{Discussion}

 \subsection{Variation of $I_{3.3}$ with $G_0$}

  The dependence of the 3.3\mum band intensity on $G_0$ results from three effects:
 change in the abundance of its carriers, PAH ionization since the 3.3\mum band is expected to be smaller for PAH$^+$ relative to PAH$^0$ \citep[see a review on quantum-chemical calculations by][]{pauzat11}, and excitation effects. We discuss these effects here by considering the $I_{3.3}/I_{PAH^0}$ ratio.  As shown in Fig. \ref{figure_correlations}a, this ratio  presents some dispersion, but with an overall increase with $G_0$.  

The observed intensity of the 3.3 \mum band can be written as a function of   the column density $N_{3.3}$ of its carriers  and  $G_0$: 

\begin{eqnarray} 
I_{3.3} & = &   G_0 \,  N_{3.3}\, \epsilon_{3.3}   
,\end{eqnarray} 

\noindent where $\epsilon_{3.3}$ is  the average emissivity  at 3.3\mum\  per aromatic CH bond in PAHs along the line of sight, which depends on the molecular properties of their carriers and the excitation conditions, and $N_{3.3}$ is the column density of aromatic CH bonds.

 The intensity of the 3.3 \mum band  is the sum of the contribution of the neutral and cationic PAHs.  Thus, the ratio $I_{3.3}/I_{PAH^0}$ can be written as

\begin{eqnarray}
\frac{I_{3.3}}{I_{PAH^0}} & = &  \frac{I_{3.3}^0}{I_{PAH^0}}  \left(1 + \frac{I_{3.3}^+}{I_{3.3}^0}  \right) = \frac{N_{3.3}^0}{N_{PAH^0}} \frac{\epsilon_{3.3}^0}{\epsilon_{PAH}^0}    \left(1 + \frac{N_{3.3}^+}{N_{3.3}^0}  \frac{\epsilon_{3.3}^+}{\epsilon_{3.3}^0} \right)
,\end{eqnarray}

\noindent  where $I_{3.3}^{0/+}$, $\epsilon_{3.3}^{0/+}$  are the emitted intensity and average emissivity of the 3.3\mum band due to \pah and \pahp, and $N_{3.3}^{0/+}$ is the column density of aromatic CH bonds of neutral and ionized species. 
We  assume that the ionization fraction is the same for the carriers of the 3.3\mum band and for the whole PAH population, that is,{\it } $N^+_{3.3}/N^0_{3.3} = N_{PAH^+}/N_{PAH^0}$  \citep[][see discussion below]{montillaud13}. From this we obtain 

\begin{eqnarray} 
\frac{I_{3.3}}{I_{PAH^0}} =  k_{3.3} \left(1 + \frac{N_{PAH^+}}{N_{PAH^0}} p_{3.3}  \right)
\label{eqnc}
,\end{eqnarray}

\noindent where we have defined 

\begin{eqnarray}
k_{3.3} &=& \frac{\epsilon_{3.3}^0}{\epsilon_{PAH}^0}   \frac{N_{3.3}^0}{N_{PAH^0}} \\
p_{3.3}&=&\frac{\epsilon_{3.3}^+}{\epsilon_{3.3}^0}.
\end{eqnarray}

$k_{3.3}$ represents the fraction of  the bolometric PAH emission in the 3.3\mum band for the \pah population and is therefore an indicator of the PAH size distribution.  \citet{pech02}  have modeled the IR emission spectrum of a distribution of PAHs under irradiation by UV photons and derived values for $k_{3.3}$ of about $ 0.01-0.02$  depending on the assumed size distribution.
The $p_{3.3}$ parameter represents the  emissivity ratio for PAH cations and neutrals. It is expected to depend on the excitation conditions since PAH cations (open-shell species) have more electronic bands at low energy. However, in a relatively hard UV field, the excitation will be dominated by the strong UV $\pi\rightarrow\pi^*$ and $\sigma\rightarrow\sigma^*$ type bands, which are present in all charge states and are almost equal across them \citep{mulas06}. The $p_{3.3}$ parameter will then depend mainly on the relative IR absorption cross-section at 3.3\mum of PAH cations relative to neutrals. Values  in the range 0.2-0.5 have been calculated for PAH sizes between  20 and 60 C atoms \citep{malloci07}.  

The total emissivity for a neutral and a cation with similar size is approximately the same because of conservation of energy and the fact that their UV absorption cross-sections are in first approximation similar, as explained above. Therefore, we can use the ratio of the observed intensity $I_{PAH^+}/I_{PAH^0}$ as a probe for the column density ratio of ionized to neutral PAHs, and rewrite Eq. \ref{eqnc} using only the observed intensities and the molecular parameters $k_{3.3}$ and $p_{3.3}$:
 
\begin{eqnarray} 
\frac{I_{3.3}}{I_{PAH^0}}  =  k_{3.3} \left(1 + \frac{I_{PAH^+}}{I_{PAH^0}}\,p_{3.3}  \right). 
\label{eqkp}
\end{eqnarray} 

The term in parenthesis is related to the contribution of the ionization fraction of PAHs along the line of sight to the observed $I_{3.3}/I_{PAH^0}$ ratio.  Using the maps we obtained with PAHTAT, we derived the values of  $G_0$ and $I_{PAH^+}/I_{PAH^0}$ for each pixel  (Figs. \ref{fig_g0} and \ref{figure_pahtat}f, respectively). Then, we used these values to compute a  model map of $I_{3.3}/I_{PAH^0}$ using Eq. \ref{eqkp} with constant values of $k_{3.3}$ and $p_{3.3}$. 
This semi-empirical model (hereafter model 1, Fig. \ref{figure_correlations}b)  cannot reproduce the observed trend in  $I_{3.3}/I_{PAH^0}$.  As an example, Fig.\,\ref{figure_correlations}b shows the results for model 1 using $k_{3.3} = 0.2$ and $p_{3.3} = 0.5$.  These values  give the best fit to the data within the physically meaningful range for $k_{3.3}$ and $p_{3.3}$. 
This shows that the observed variations of  $I_{3.3}/I_{PAH^0}$  cannot be accounted for by a variation in the mean charge of PAHs alone. 

To obtain a better fit of the observed increase of $I_{3.3}/I_{PAH^0}$ with $G_0$, we assumed that   $k_{3.3}$ depends on $G_0$ with a linear relationship: $k_{3.3}= 0.02 + k_{3.3}^{*} \times G_0 $.  We fit the slope of this correlation by varying $k_{3.3}^{*}$. The best fit is obtained using $k_{3.3}^{*} = \ex{8}{-7}$ (hereafter model 2, Fig. \ref{figure_correlations}c). 
The dependency of $k_{3.3}$ on $G_0$ could be due to  an increase of the  abundance of the 3.3\mum carrier  relative to all neutral PAHs, or to an increase of the emissivity at 3.3 \mum\, relative to the total emissivity of \pah. As discussed below, the first effect is likely to involve a chemical evolution of the PAH population, whereas the second is related to excitation conditions.

 Analyzing infrared observations obtained by Spitzer and Herschel,  \citet{berne12} showed that there is no global variation of the PAH abundance in the region covered by our observations.
\citet{montillaud13}  modeled the chemical evolution  of PAHs in NGC 7023 NW. They showed that PAHs of small sizes (with a number of carbon atoms $N_C \lesssim 50$) are fully dehydrogenated, whereas larger PAHs are normally hydrogenated and possibly superhydrogenated for the largest sizes ($N_C \gtrsim 90$).  The ionization fraction is found to be rather similar from one size to the other. It is interesting to note that across the PDR,  intermediate-sized
PAHs ($N_C \sim 60-70$) can experience strong variations in their hydrogenation state because of the competition between photodissociation by UV photons and recombination with H atoms. For instance, C$_{66}$H$_{20}$ is predicted to be fully hydrogenated at the border of the PDR, but fully dehydrogenated deeper inside the cloud at a distance from 5 to 10\arcsec\, from the PDR front. Here, the abundance of H relative to H$_2$ decreases faster than the attenuation of UV photons, resulting in fully dehydrogenated species.  This effect could explain the  increase in the  3.3\mum band intensity  with increasing $G_0$.
Another possibility is that the emissivity at 3.3\mum   increases with $G_0$ as a result of the higher average internal temperatures of PAHs with higher UV field. Two effects may be invoked: in the most exposed layers of the PDR, the UV radiation field is becoming harder because of decreased extinction by dust \citep{rapacioli06}. Secondly, when the UV flux increases, the probability for multiple photon events increases. A PAH that has absorbed a UV photon can only relax part of its internal energy before absorbing another photon, and it is therefore heated to higher temperatures. However, whereas multiple photon events are important for studying photo-dissociation \citep{montillaud13}, they remain rare events in NGC 7023 NW and  do not contribute significantly to the mid-IR emission. 
In addition, \citet{witt06} showed that in the first layers of the PDR there is a significant contribution of high-energy UV-photons ($h\nu \gtrsim 10.3$ eV) that are suppressed in the more protected layers ($A_V \gtrsim 2$). 
The variation of the hardness of the UV field across the PDR provides another explanation for the increase of the intensity of the 3.3\mum band with $G_0$. Quantifying the importance of this process compared to the hydrogenation effect described above would require a full chemical and photo-physical model, which is not the purpose of this paper. We conclude  that the increase of $I_{3.3}/I_{PAH^0}$ with $G_0$ is due to a change in the hardness of the UV field combined with an increased abundance of aromatic CH bonds at the border of the PDR, both effects being relatively moderate. 

\begin{figure*}
\centering
\begin{subfigure}{.5\textwidth}
  \centering
  \includegraphics[width =\textwidth]{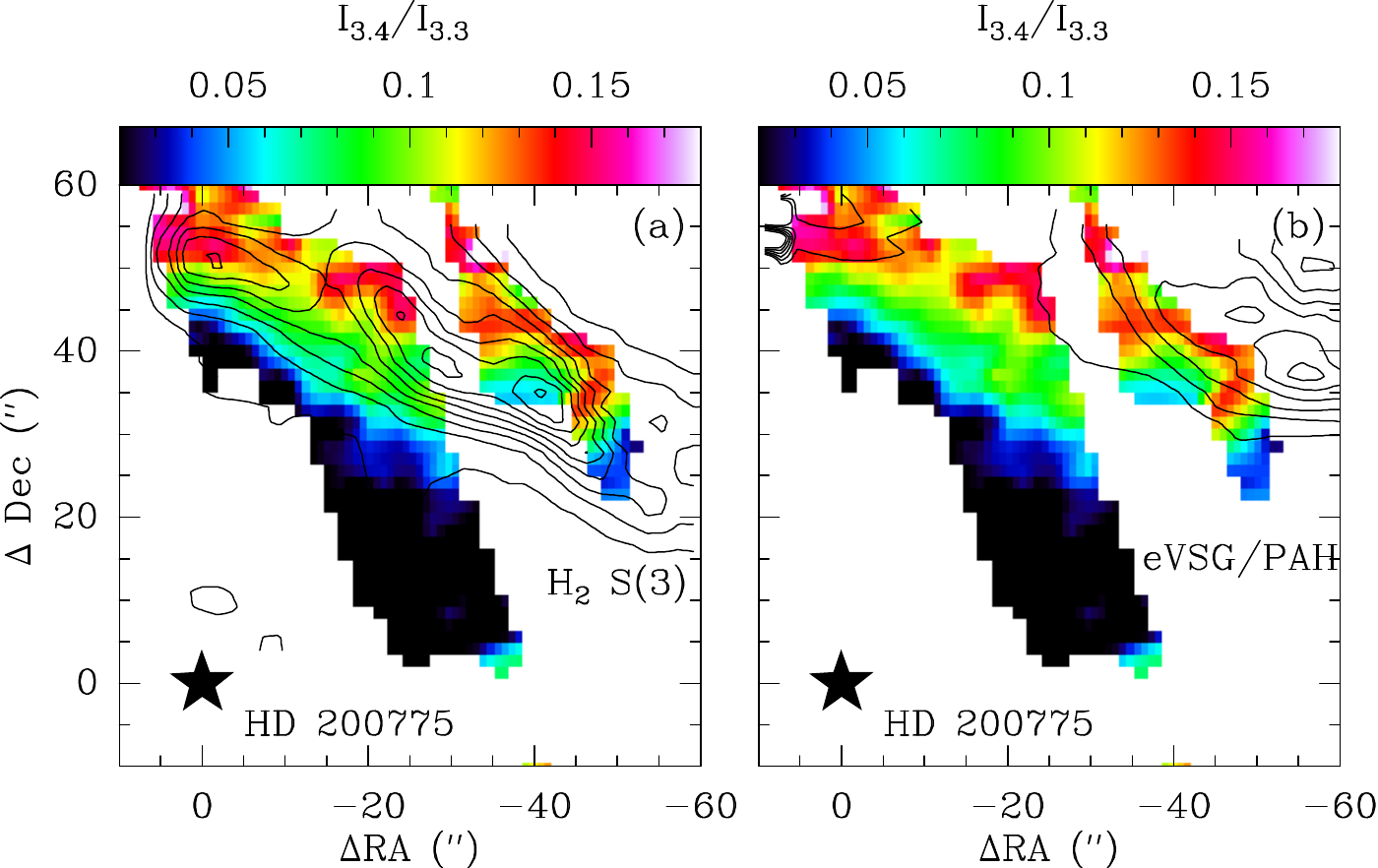}
  \label{fig:sub1}
\end{subfigure}%
\begin{subfigure}{.4\textwidth}
  \centering
\includegraphics[angle = -90, trim = 0cm 2cm 0cm 0cm, clip, width =\textwidth]{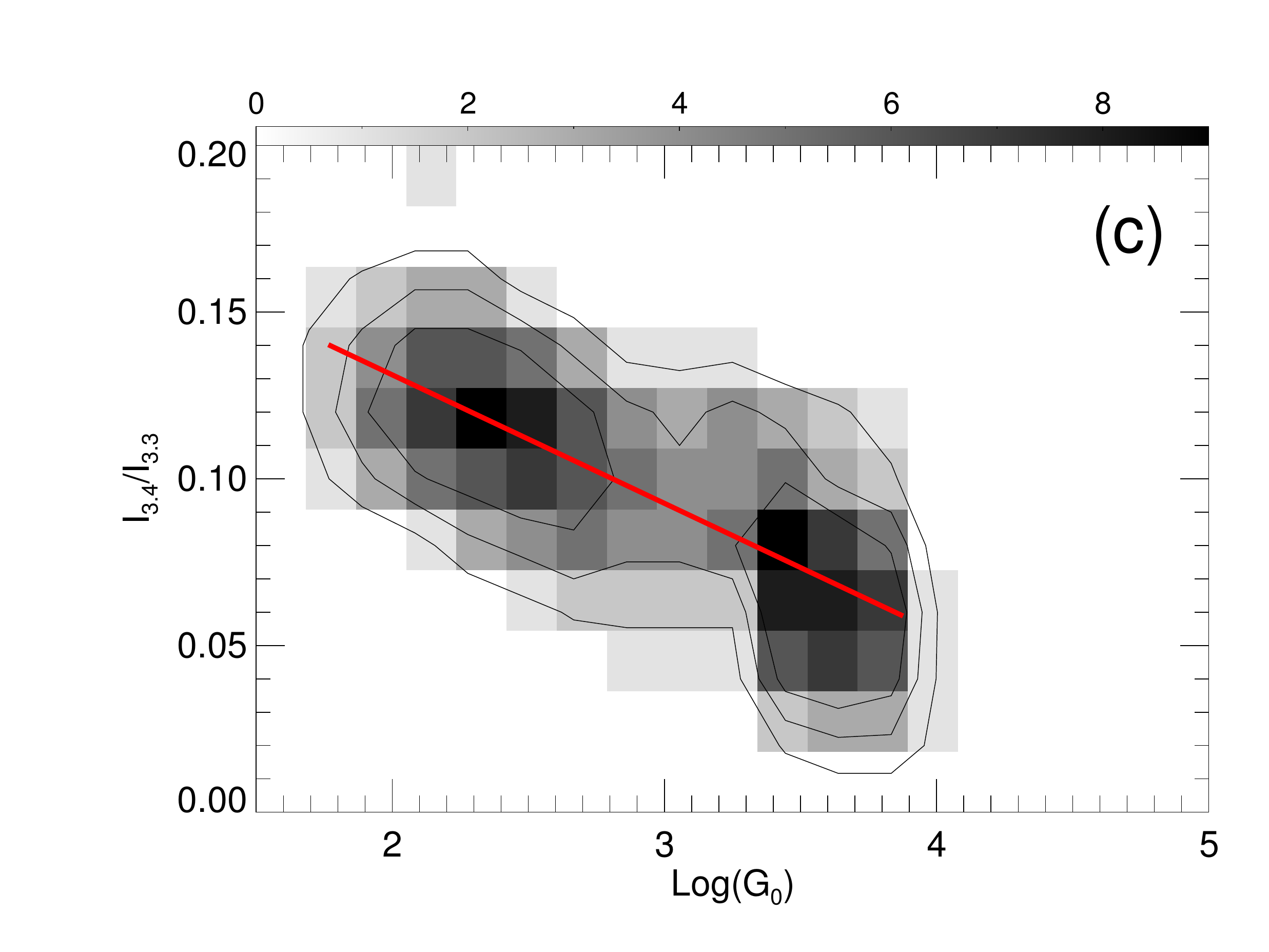}  
  \label{fig:sub2}
\end{subfigure}
\caption{ Map of the 3.4/3.3 \mum ratio (color scale) with the contours of the H$_2$ S(3) line ({\it a}), and of  the ratio $I_{eVSG}/I_{PAH} = I_{eVSG}/(I_{PAH^0}+I_{PAH^+}) $ ({\it b}). The density plot of  the $I_{3.4}/I_{3.3}$ band ratios vs $G_0$ is also shown ({\it c}). In red, we report the linear regression of the data points. 
Only the pixels for which  $I_{3.3} > \ex{1.0}{-7}$\wmqsr have been used to build in the analysis of the 3.4/3.3 \mum ratio.}
\label{figure_akari}
\end{figure*}

 \subsection{Photo-destruction of aliphatics}
Similar to the 3.3\mum band,  the intensity of the 3.4\mum band relative to the total intensity of the \pah bands ($I_{3.4}/I_{PAH^0}$) also presents some dispersion. However,  it shows an overall decrease with $G_0$  (Fig. \ref{figure_correlations}d). 
In analogy with the 3.3\mum case, the observed intensity of the 3.4\mum band can be written as 
\begin{eqnarray} 
I_{3.4} & = &   G_0 \,  N_{3.4}\, \epsilon_{3.4}   
,\end{eqnarray} 
\noindent where $\epsilon_{3.4}$ and $N_{3.4}$ are the average emissivity  and column density per aliphatic CH bond along the line of sight. Equation \ref{eqkp} can be rewritten for the 3.4\mum band as 

\begin{eqnarray} 
\frac{I_{3.4}}{I_{PAH^0}} =  k_{3.4}  \times \left(1 + \frac{I_{PAH^+}}{I_{PAH^0}} \times p_{3.4}  \right)
\label{eqfinalalip}
,\end{eqnarray} 

\noindent where we have defined

\begin{eqnarray}
k_{3.4} &=& \frac{\epsilon_{3.4}^0}{\epsilon_{PAH}^0}   \frac{N_{3.4}^0}{N_{PAH^0}}  = \frac{I_{3.4}^0}{I_{PAH}^0} \label{eq34}\\
p_{3.4}&=&\frac{\epsilon_{3.4}^+}{\epsilon_{3.4}^0}.
\end{eqnarray}

To obtain an estimate of the typical values for $p_{3.4}$, we used the NASA-Ames PAH IR Spectroscopic database\footnote{\url{http://www.astrochem.org/pahdb/}} \citep[hereafter the NASA-Ames PAH database,][]{boersma14} and obtained $p_{3.4} \gtrsim 1$, depending on the size of the PAH being considered.  Since $p_{3.4} > p_{3.3}$, the term in parenthesis is larger for the 3.4\mum than for the 3.3\mum band, implying that the charge state of PAHs modifies the ratio $I_{3.4}/I_{PAH^0}$ more significantly than does $I_{3.3}/I_{PAH^0}$. 

Assuming $p_{3.4} = 1$, the best fit to the data using our model 1 is obtained using  $k_{3.4}= 0.002$, which allows us to reproduce  the average values of  $I_{3.4}/I_{PAH^0}$ (Fig. \ref{figure_correlations}e). $k_{3.4}$ is found to be a factor of 10 lower than $k_{3.3}$  because of the lower abundance of aliphatic CH groups compared to the aromatic CH groups.  The relative abundance of aliphatic vs aromatic groups we derive agrees with previous studies, including the recent work of \citet{li12}. The observed   negative slope of the $I_{3.4}/I_{PAH^0}$ vs $G_0$ correlation    (Fig.   \ref{figure_correlations}d) is  not reproduced by our model 1 (Eq. \ref{eqfinalalip} and Fig.   \ref{figure_correlations}e). 
As for the 3.3\mum band, we assume a dependency of $k_{3.4}$ with $G_0$: $k_{3.4}= 0.002 + k_{3.4}^{*} \times G_0 $.
 The observations can be best reproduced by $k_{3.4}^* = -  \ex{1.8}{-7}$ (Fig. \ref{figure_correlations}e).  This negative term mainly reflects the destruction of the carriers of the 3.4\mum band with increasing $G_0$. This is an effective value that reflects both the increase in intensity and hardness of the UV field while reaching the edge of the PDR.

\subsection{Variation of $I_{3.4}/I_{3.3}$ with $G_0$}

The ratio 
$I_{3.4}/I_{3.3}$ can be used as a tracer of the aliphatic to aromatic content in PAHs if we assume that both bands arise from the same carriers. 
This ratio is low in the cavity  ($\sim 0.03$ at P1) and increases monotonically  toward the PDR, reaching its highest value ($\sim 0.15$) slightly behind the H$_2$ S(3) and the \pah emission (Fig. \ref{figure_akari}a).  The highest gradient is found at the PDR front where in only $\sim10\arcsec $ the ratio increases by a factor of $\sim 3$.  This shows that photo-chemical processing of the carriers of the 3.4\mum band is  taking place in this region. We note that $10\arcsec$ corresponds to the angular  width of the filament  traced by the H$_2$ rotational emission.
Interestingly, this is also the evaporation zone of eVSGs into gas-phase PAHs. 
We have used the $I_{eVSG}/I_{PAH} = I_{eVSG}/(I_{PAH^0}+I_{PAH^+}) $ ratio as an indicator of the destruction of eVSGs into PAHs by UV photons. The relative spatial distributions suggest that the destruction of eVSGs is then followed by that of the 3.4\mum carriers. This suggests that the evaporation of eVSGs leads to the production of PAHs with aliphatic groups.

 Figure \ref{figure_akari}c shows the pixel-to-pixel correlation between $I_{3.4}/I_{3.3}$ and $\log(G_0)$. 
  Dividing each side of Eq. (8) by Eq.  (6), we obtain 
$$\frac{I_{3.4}}{I_{3.3}} =  \frac{N_{3.4}}{N_{3.3}}  \frac{\epsilon_{3.4}^0}{\epsilon_{3.3}^0} D_{ion,}$$

\noindent where $D_{ion}$ is the ratio of the terms in parenthesis in each equation and represents the variation of $I_{3.4}/I_{3.3}$ due to ionization. Because the intensity of \pahp relative to that of \pah 
is $\lesssim 0.4$, $D_{ion}$ is very close to 1. More specifically, $D_{ion}$ is about 10\% higher in the more exposed regions in which $I_{PAH^+}/I_{PAH^0} = 0.4$ compared to the region with no \pahp emission, thus it cannot explain the trend shown in Fig. \ref{figure_akari}c. For PAH sizes larger than $\sim50$ carbon atoms, $\epsilon^0_{3.4}/\epsilon^0_{3.3}$ can be also considered constant (cf. NASA-Ames PAH database), and  
therefore the observed trend can be mainly attributed to a decrease in the abundance of the aliphatic CH bonds relative to aromatic CH bonds. 

 The trend we observe is similar to that reported in \citet{joblin96} and analyzed by the authors using a simple photo-chemical model that describes the evolution of methylated PAHs.  The authors showed that the decrease of the 3.4\mum band occurs because the reactions that could reconstruct the aliphatic side-groups in the gas-phase are highly improbable.

\section{Conclusions}

Thanks to the high spatial resolution of observations made in the near- and mid-IR,  we probed the evolution of aliphatic and aromatic CH groups in the spatially resolved PDR NGC 7023 NW.  We showed that the intensity of the 3.3\mum band relative to the total PAH emission increases with $G_0$, while the relative contribution of the 3.4\mum band decreases with $G_0$.  In the most exposed layers of the PDR, the UV radiation field is harder and high-energy UV photons can excite PAHs to higher temperatures, leading thus to an increase of the fraction of energy emitted at 3.3\mum. The higher abundance of H atoms in this region can also increase the abundance of CH aromatic bonds. On the other hand, the higher flux of UV photons leads to efficient destruction of the  more fragile aliphatic CH bonds attached to PAHs, which cannot be efficiently reformed in these regions. In addition, we  showed that the observed 3.45\mum plateau is dominated by the emission from aromatic bonds.

We find that the change in the aliphatic vs aromatic composition is particularly important along the filament that delineates the PDR. In particular, the $I_{3.4}/I_{3.3}$ ratio peaks close to the region where eVSGs are photo-evaporated into PAHs. 
This suggests that the processing of eVSGs leads to PAHs with attached aliphatic sidegroups, which provides further insights into the formation and evolution of these species in astrophysical environments. 
 (Very) small grains of mixed aromatic  and aliphatic composition are known to be present in circumstellar environments \citep{kwok01, goto03, sloan07}, and they have been considered to be a major component in the dust model by \citet{jones12a}. We provided here the first observational insights for the presence of such grains in interstellar clouds.

\begin{acknowledgements}
 We thank the anonymous referee for  useful comments. This work was supported by the
French National Program, Physique et Chimie du Milieu
Interstellaire, which is gratefully acknowledged.\\
P. Pilleri acknowledges financial support from the Centre National d'Etudes Spatiales (CNES).\\ The research leading to these results has also received funding from the European Research Council under the European Union's Seventh Framework
Programme (FP/2007-2013)  ERC-2013-SyG, Grant Agreement n. 610256 NANOCOSMOS. \\
\end{acknowledgements}

\bibliographystyle{aa}
\bibliography{biblio}

\end{document}